\documentclass{article}
\usepackage{spconf,amsmath,graphicx}
\usepackage{amsmath} 
\usepackage{graphicx}
\usepackage{caption}
\usepackage{subfig}
\usepackage{lipsum}
\usepackage{multirow,booktabs}
\usepackage{multicol}
\usepackage{tabularx}

\usepackage{enumitem}
\usepackage{url}
\usepackage{hyperref}

\title{Generating Adaptive and Robust Filter Sets using an Unsupervised Learning Framework}
\name{Mohit Prabhushankar, Dogancan Temel, and Ghassan AlRegib} 
\address{Center for Signal and Information Processing, School
	of Electrical and Computer Engineering, \\ Georgia Institute of Technology, Atlanta, GA}
\begin{document}

\onecolumn 

\begin{description}[labelindent=1cm,leftmargin=3cm,style=multiline]

\item[\textbf{Citation}]{M. Prabhushankar, D. Temel and G. AlRegib, "Generating adaptive and robust filter sets using an unsupervised learning framework," 2017 IEEE International Conference on Image Processing (ICIP), Beijing, 2017, pp. 3041-3045.} \\

\item[\textbf{DOI}]{\url{https://doi.org/10.1109/ICIP.2017.8296841}} \\

\item[\textbf{Review}]{Date added to IEEE Xplore: 22 February 2018} \\

\item[\textbf{Code/Poster}]{\url{https://ghassanalregib.com/publications/}} \\

\item[\textbf{Bib}] {
@INPROCEEDINGS\{Temel2017\_ICIP,\\ 
author=\{M. Prabhushankar and D. Temel and G. AlRegib\},\\ 
booktitle=\{2017 IEEE International Conference on Image Processing (ICIP)\},\\ 
title=\{Generating adaptive and robust filter sets using an unsupervised learning framework\},\\ 
year=\{2017\},\\
pages=\{3041-3045\},\\
doi=\{10.1109/ICIP.2017.8296841\},\\ 
ISSN=\{2381-8549\},\\ 
month=\{Sept\},\}\\
} \\

\item[\textbf{Copyright}]{\textcopyright 2017 IEEE. Personal use of this material is permitted. Permission from IEEE must be obtained for all other uses, in any current or future media, including reprinting/republishing this material for advertising or promotional purposes,
creating new collective works, for resale or redistribution to servers or lists, or reuse of any copyrighted component
of this work in other works. } \\

\item[\textbf{Contact}]{\href{mailto:alregib@gatech.edu}{alregib@gatech.edu}~~~~~~~\url{https://ghassanalregib.com/}\\ \href{mailto:dcantemel@gmail.com}{dcantemel@gmail.com}~~~~~~~\url{http://cantemel.com/}}
\end{description} 

\thispagestyle{empty}
\newpage
\clearpage

\twocolumn

\ninept
\maketitle
\begin{abstract}

In this paper, we introduce an adaptive unsupervised learning framework, which utilizes natural images to train filter sets. The applicability of these filter sets is demonstrated by evaluating their performance in two contrasting applications - image quality assessment and texture retrieval. While assessing image quality, the filters need to capture perceptual differences based on dissimilarities between a reference image and its distorted version. In texture retrieval, the filters need to assess similarity between texture images to retrieve closest matching textures. Based on experiments, we show that the filter responses span a set in which a monotonicity-based metric can measure both the perceptual dissimilarity of natural images and the similarity of texture images. In addition, we corrupt the images in the test set and demonstrate that the proposed method leads to robust and reliable retrieval performance compared to existing methods. 
\end{abstract}
\vspace{-1.5mm}
\section{Introduction}
\label{sec:intro}
\vspace{-1.5mm}
Unsupervised learning is a branch of machine learning tasked with inferring functions that describe hidden structures from unlabeled data. In terms of neural networks, these functions are a set of affine transformations subjected to nonlinearity. Each function is described as a layer, and stacking layers leads to deep neural nets. These deep networks can be used as end to end systems utilizing all the underlying layers as one entity. An example of vanilla neural networks are autoencoders. The authors in \cite{Goodfellow-et-al-2016} describe autoencoders as neural networks that are trained to copy inputs to outputs. By placing multiple constraints, the network learns to extract hidden characteristics from data. Autoencoders can extract overcomplete basis functions which can be used for denoising. Stacking layers of denoising autoencoders, which are trained layer-wise to locally denoise corrupted versions of their inputs, led to developement of Stacked Denoising Autoencoders \cite{vincent2010stacked}. 

In this article, we show that the individual layers can be repurposed to tasks different from what they were trained for. In other words, we use neural networks as tools to generate affine functions. Mathematically, these functions act as filters that process inputs to span a response set in which we can perform various image processing tasks. To test the applicability of generated filters, we focus on image quality assessment and texture retrieval. These applications are a challenge for the traditional deep learning networks due to lack of large labeled datasets that are essential for training. To overcome label requirements and application specific data, we learn filters from natural images in an unsupervised fashion. In practice, we need to train filters that are insensitive to lower order statistics of natural images, so that the trained model can work on disparately distributed images during testing. We achieve this by proposing a simple extension to ZCA whitening procedure during preprocessing in Section \ref{SubSec:Preprocessing}, and describe the learning architecture that generates the filter sets in Section \ref{SubSec:Linear Decoder}. We integrate the preprocessing and the learning blocks to obtain the Unsupervised Learning Framework which has two tuning parameters. These parameters can be adjusted to generate multiple filter sets capturing multi-order statistics as discussed in Section \ref{sec:Filter sets}. We utilize the generated filters to perform image quality assessment in Section \ref{sec:IQA} and texture retrieval in Section \ref{sec:Texture}. Finally, we conclude our work in Section \ref{sec:Conclusion}.
\vspace{-1.5mm}
\section{Unsupervised Learning Framework}
\label{sec:ULF}
\begin{figure}[htbp!]
	\begin{center}
		\noindent
		\includegraphics[width=1\linewidth]{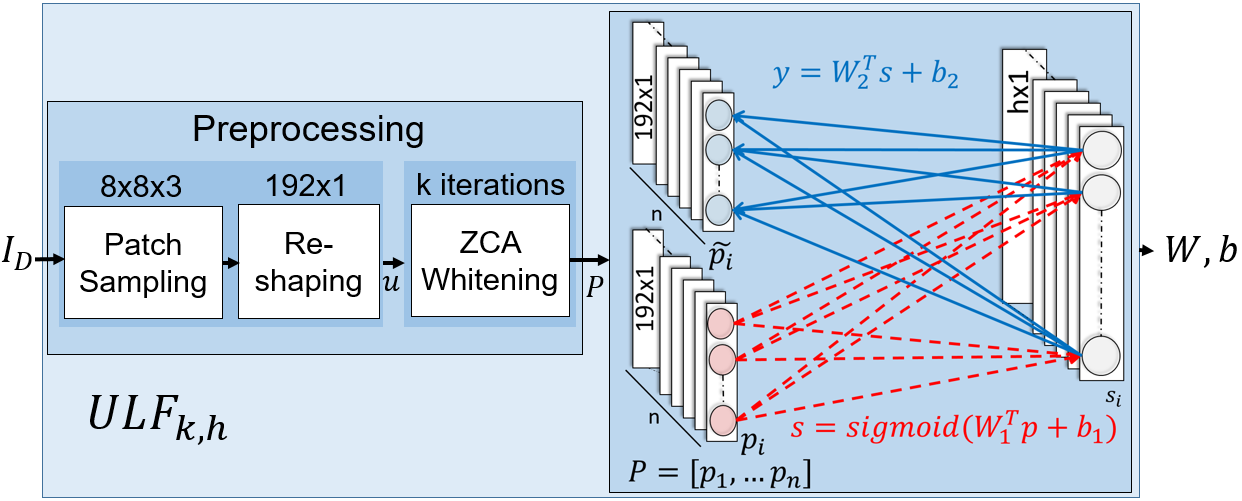}
		\caption{Unsupervised Learning Framework (ULF).}
		\label{fig:framework}
	\end{center}
	\vspace{-5.0mm}
\end{figure}

\vspace{-1.5mm}
In this section, we describe the Unsupervised Learning Framework $(ULF_{k,h})$, whose block diagram is shown in Fig. \ref{fig:framework}. The framework takes in an input $I_D$ and generates weights $W$ and bias $b$, which span our filter sets. The subscripts $k$ and $h$ correspond to tuning parameters that determine filter characteristics.
\vspace{-5.0mm}
\subsection{Preprocessing}
\label{SubSec:Preprocessing}
As shown in Fig. \ref{fig:framework}, the preprocessing stage converts raw images from input database $I_{D}$ into patch-based vectors $u$, and feeds them into the Zero-Phase Component Analysis (ZCA) whitening algorithm \cite{bell1997independent}. ZCA whitening is a decorrelating procedure that removes the first and the second order statistics of input data and forces the ensuing learning stage to capture higher order statistics. Essentially, this specifies the covariance matrix of output $P$ to be the identity matrix $I$. Therefore, the sample covariance matrix of $P$ is given by
\begin{equation}\label{PPTransposeIdentity}
PP^{T} = (n-1)I,
\end{equation} 
where $n$ is the number of input vectors $u$. For a large value of $n$, the sample covariance estimate used in Eq. \ref{PPTransposeIdentity}, which is the maximum likelihood estimate (MLE), would be close to the actual covariance matrix. ZCA whitening attempts to find a set of linear, symmetric filters $W_z$ that transform $u$ to $P$. The output of the preprocessing block is given by
\begin{equation}\label{ZCATransform}
P = W_{z}u,
\end{equation} 
where $u$ is assumed to be zero centered. Following the approach described by the authors in \cite{bell1997independent} and \cite{krizhevsky2009learning}, we calculate $W_z$ as
\begin{equation}\label{ZCA Filters}
W_{z} \propto (uu^T)^{-1/2}.
\end{equation} 
The outer product $uu^T$ is symmetric and orthogonally diagonalizable. Hence, applying eigenvalue decomposition to $uu^T$, we obtain
\begin{equation}\label{ZCA Filters Diagnolized}
uu^T = V\Lambda V^T,
\end{equation}
where $V$ are the eigenvectors and $\Lambda$ are the eigenvalues of $uu^T$. Raising both sides of Eq. \ref{ZCA Filters Diagnolized} to the power $-1/2$, we get
\begin{equation}
\label{ZCA Diagnolized power}
(uu^T)^{-1/2} = V\Lambda^{-1/2}V^T.
\end{equation}
Combining Eq. \ref{ZCA Diagnolized power} and Eq. \ref{ZCA Filters}, we obtain an expression for $W_z$ as
\begin{equation}
W_z \propto V\Lambda^{-1/2}V^T.
\end{equation} 
The matrix $uu^T$ is, however, ill conditioned. Therefore, a regularization term $\epsilon$ is added to the eigenvalues, so that the inverse of the smaller eigenvalues do not arbitrarily scale the resultant. Hence, the expression for the filters of a standard ZCA whitening algorithm is given by 
\begin{equation}\label{ZCA Filters in practice}
W_z \propto V(\Lambda+\epsilon)^{-1/2}V^T.
\end{equation}
Applying $W_z$ on $u$ produces whitened patches whose covariance matrix is ideally $I$. In practice, this is not always the case. Fig. \ref{fig:Visualization Covariane Matrices}(a) shows $uu^T$ while Fig. \ref{fig:Visualization Covariane Matrices}(b) shows $PP^T$ using Eq. \ref{ZCATransform} and Eq. \ref{ZCA Filters}. The nine distinct blocks are present in Fig. \ref{fig:Visualization Covariane Matrices}(a) because of the three separate channels (RGB) whose distributions are different but not entirely uncorrelated. These blocks are still present in Fig. \ref{fig:Visualization Covariane Matrices}(b) because of the approximation caused by the addition of $\epsilon$. Such an approximation is tolerable as long as the testing data distribution is similar to the training data distribution. Since we propose utilizing texture images for testing, which do not adhere to the natural scene statistics, an approximate identity matrix for $PP^T$ is not acceptable. Hence, we propose a simple extension to the standard ZCA algorithm in which we apply Eq. \ref{ZCATransform} iteratively on the initial input $(u_0)$, $k$ times. k is specified as an input to the algorithm. Using this extension, the intermediate whitened vector denoted by $u_i$, at every $i^{th}$ iteration is given by
\begin{equation}\label{ZCATransform Extension}
\begin{gathered}
u_i = W_z^{i}u_{i-1}, \forall i \in [1 : k], \forall k > 0\\
W_z^i \propto (u_{i-1}u_{i-1}^T)^{-1/2}, \\
\end{gathered}
\end{equation}
where $W_z^i$ is the intermediate whitening matrix at iteration $i$. As $u_i$ can be completely described with the knowledge of $u_{i-1}$, the sequence of matrices $u_i, \forall i = [1,k]$ satisfy the markov property. Hence after calculating all the $W_z^i$ matrices using Eq. \ref{ZCATransform Extension}, we express the final whitened patches P as
\begin{equation}\label{ZCATransform Extension Final}
\begin{gathered}
P = W_{z}^k u_{k-1},\\
\end{gathered}
\end{equation} 
where $W_z^k$ and $u_{k-1}$ are the whitening filters, and input vectors at $k^{th}$ iteration. $k = 1$ gives us the standard ZCA algorithm. Taking $k$ to be $1/\epsilon$, with $\epsilon = 10$, we obtain the covariance matrix as shown in Fig. \ref{fig:Visualization Covariane Matrices}(c). Essentially, by performing whitening multiple times, the effect of $\epsilon$ is reduced. This ensures that Eq. \ref{PPTransposeIdentity} holds and the lower order statistics from natural images is eliminated.

In common practice, the whitening filters $(W_z)$ calculated during training are used to transform the test data as well, because there is not enough test data ($n$) to estimate sample covariance matrix in Eq. \ref{PPTransposeIdentity}. However, $W_z$ is trained on natural images and is not suitable to transform texture or distorted images. Hence, we calculate $W_z$ during testing as well. Even though $n$ during testing is insufficient to estimate covariance matrix precisely, it is not singular and for the present we overlook this approximation.
\subsection{Linear Decoder} 
\label{SubSec:Linear Decoder}
Linear decoder is a variation of standard autoencoder in which sigmoid nonlinearity of the final reconstructed output is replaced with a linear activation function. The primary task of this network is to learn representative structure in its input data while attempting to reconstruct it. Adding a sparsity criterion forces the network to learn unique statistical features from the data rather than an identity function in the hidden layer \cite{Goodfellow-et-al-2016}. The hidden layer responses $(s)$, with a sigmoid nonlinearity activation, are obtained as
\begin{equation}\label{ForwardFilters}
\begin{gathered}
s = sigmoid(W_1^T P + b_1),\\
\forall  s\in \Re^{h\times n}, W_1\in \Re^{d\times h}, P\in \Re^{d\times n}, b_1\in \Re^h,
\end{gathered}
\end{equation} 
where $W_1$ and $b_1$ are the forward weights and bias, and $P$ is the input patch matrix. Eq. \ref{ForwardFilters} is an affine function followed by a non linearity. The rows of weights $W_1^T$ represent a set of $h$ filters of dimension $d$ that linearly transform $P$. If $h$ is greater than $d$, we obtain an overcomplete basis which, coupled with sparsity, learns localized features \cite{Goodfellow-et-al-2016}. If $h$ is lower than $d$, then we have undercomplete basis set in which the network is forced to learn the most salient features. In this work, we use $h$ as a tuning parameter of the learning framework. 

The responses $s$ are used to reconstruct the patches $\tilde{P}$, using a set of backward weights $W_2$, and bias, $b_2$, as
\begin{equation}\label{BackwardFilters}
\tilde{P} = W_2^T s + b_2.
\end{equation}
Backpropagation is used to train weights $(W)$ and bias $(b)$ by setting up the objective function $J(W,b)$ as
\begin{equation}\label{Backpropagation}
J(W,b) = \lVert \tilde{P} - P \rVert_2^2 + \beta \sum_{j=1}^{n} {\rm KL}(\rho || \hat\rho_j) + \lambda\lVert W \rVert_2^2,
\end{equation}
where the first term is the reconstructed $L2$ norm error, the second term is the sparsity constraint, and the third term is the weight decay term used for regularization \cite{ng2011sparse}. KL-Divergence is used as a proxy for achieving sparsity because of its differentiability at the origin. Differentiability ensures that Eq. (\ref{Backpropagation}) is viable for L-BFGS minimization. $\rho$ is the desired sparse average activation while $\hat{\rho}$ is the actual average activation. The values for $\rho$, $\beta$, and $\lambda$ are set to $0.035$, $5$, and $3e^{-3}$ respectively as suggested by the authors in \cite{ng2012ufldl}.
\vspace{-1.5mm}
\section{Filter Set Generation}
\label{sec:Filter sets}
\begin{figure}[htbp!]
	\centering
	\subfloat[$PP^T$ with k=0]{{\fbox{\includegraphics[height = 0.3\linewidth,width=0.29\linewidth]{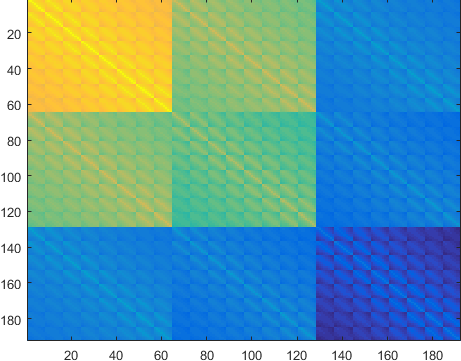} }}}%
	\subfloat[$PP^T$ with k=1]{{\fbox{\includegraphics[height = 0.3\linewidth,width=0.29\linewidth]{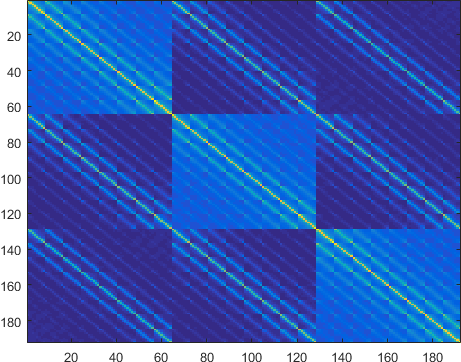} }}}%
	\subfloat[$PP^T$ with k=10]{{\fbox{\includegraphics[height = 0.3\linewidth,width=0.3\linewidth]{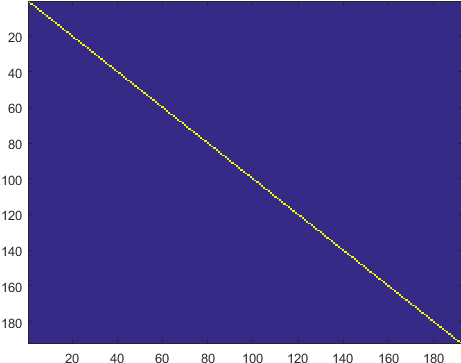} }}}%
	\vspace{-3mm}
	\subfloat[Filter sets for k = 0]{{\fbox{\includegraphics[height = 0.3\linewidth,width=0.29\linewidth]{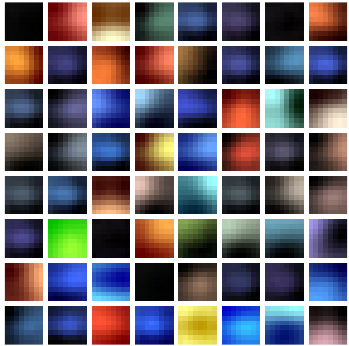} }}}%
	\subfloat[Filter sets for k = 1]{{\fbox{\includegraphics[height = 0.3\linewidth,width=0.29\linewidth]{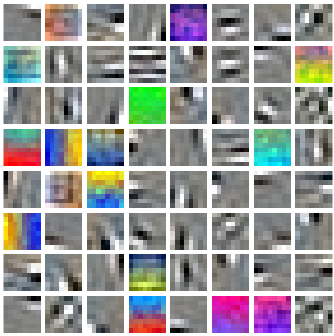} }}}%
	\subfloat[Filter sets for k = 10]{{\fbox{\includegraphics[height = 0.3\linewidth,width=0.3\linewidth]{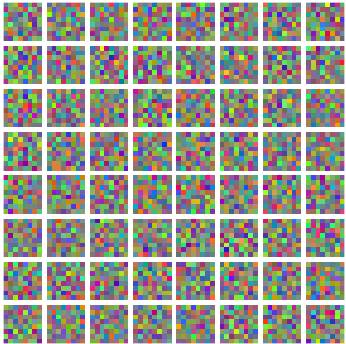} }}}%
	\caption{Covariance matrix of patches and their generated filter sets.}
	\label{fig:Visualization Covariane Matrices}
	\vspace{-5.0mm}
\end{figure}

\vspace{-1.5mm}
We use the proposed framework to generate filter sets that are trained with natural images from ImageNet database \cite{russakovsky2015imagenet}. $100$ random patches of dimensions $8\times8\times3$ are sampled from each of $1000$ random images to form $I_D$. These patches are preprocessed with different values of $k$ to train $W_1$. The weights are visualized in Fig. \ref{fig:Visualization Covariane Matrices}(d)-(f) for $h=64$. The number of neurons $(h)$, in the hidden layer, represents the number of filters for that set. The covariance matrices of patches $P$ that generate these filters are visualized in Fig. \ref{fig:Visualization Covariane Matrices}(a)-(c). Without preprocessing, filters primarily learn the first order statistics. Edges are predominant when $k=1$, which are the independent components of natural images \cite{bell1997independent}. When $k=10$, natural scenes lose pixel-wise correlations and hence there are no discernible simple structures including edges. 
\vspace{-2mm}
\section{Image Quality Assessment}
\label{sec:IQA}
\vspace{-1mm}
Perceptual image quality assessment (IQA) is a challenging field whose objective is to analyze an image and estimate its quality as perceived by humans. Database generation in this domain is a challenge, since it requires gathering subjective scores of distorted images. Learning based approaches are commonly employed in the literature to estimate quality \cite{charrier2012}-\cite{chang2011sparse}. However, all these methods are trained with distortion specific images that may not be available or sufficient. To overcome this limitation of insufficient data, we proposed an unsupervised approach to image quality assessment called UNIQUE \cite{temel2016unique} and it's extension MS-UNIQUE \cite{prabhushankar2017}, that use only generic natural images during training. In this section we recast both these estimators using the learning framework described in Section \ref{sec:ULF}. 
\vspace{-1.5mm}
\begin{figure}[htbp!]
	\begin{center}
		\noindent
		\includegraphics[width=1\linewidth]{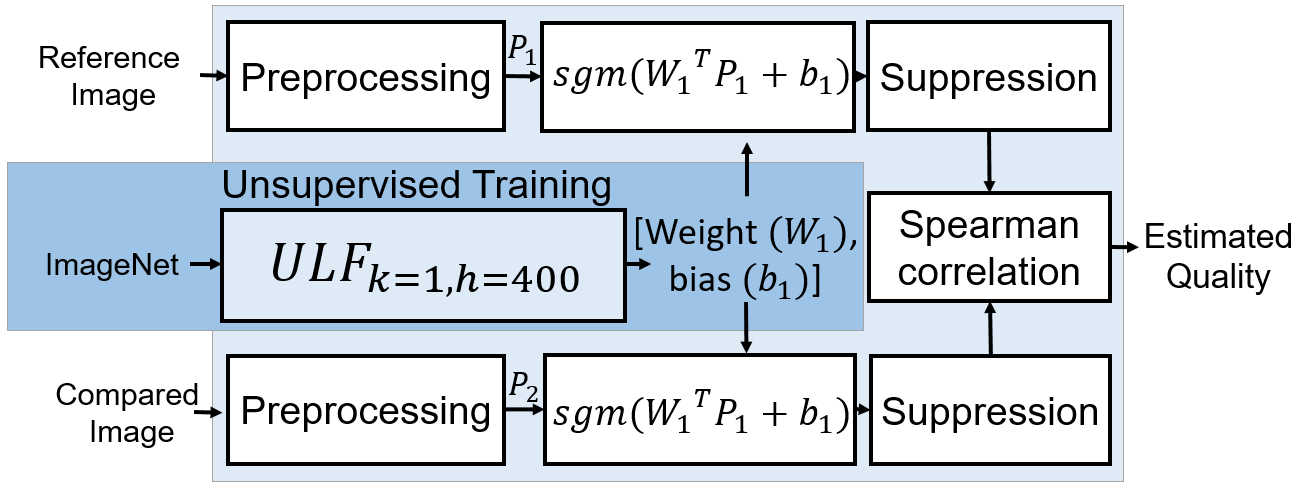}
		\caption{Unsupervised image quality estimation.}
		\label{fig:UNIQUE}
	\end{center}
	\vspace{-7.0mm}
\end{figure}
\vspace{-3mm}
\subsection{UNIQUE: Unsupervised Image Quality Estimation}
The block diagram in Fig. \ref{fig:UNIQUE} shows UNIQUE trained with the proposed framework. The proposed training uses a standard ZCA procedure with $k$ = $1$ and an overcomplete architecture with $h$ = $400$ \cite{ng2012ufldl}. The feature vectors are obtained by filtering non overlapping preprocessed patches of both reference and distorted images. The two vectors are compared using Spearman correlation to estimate quality. A complete description of the algorithm is provided in \cite{temel2016unique}.
\begin{figure}[htbp!]
	\begin{center}
		\noindent
		\includegraphics[width=1\linewidth]{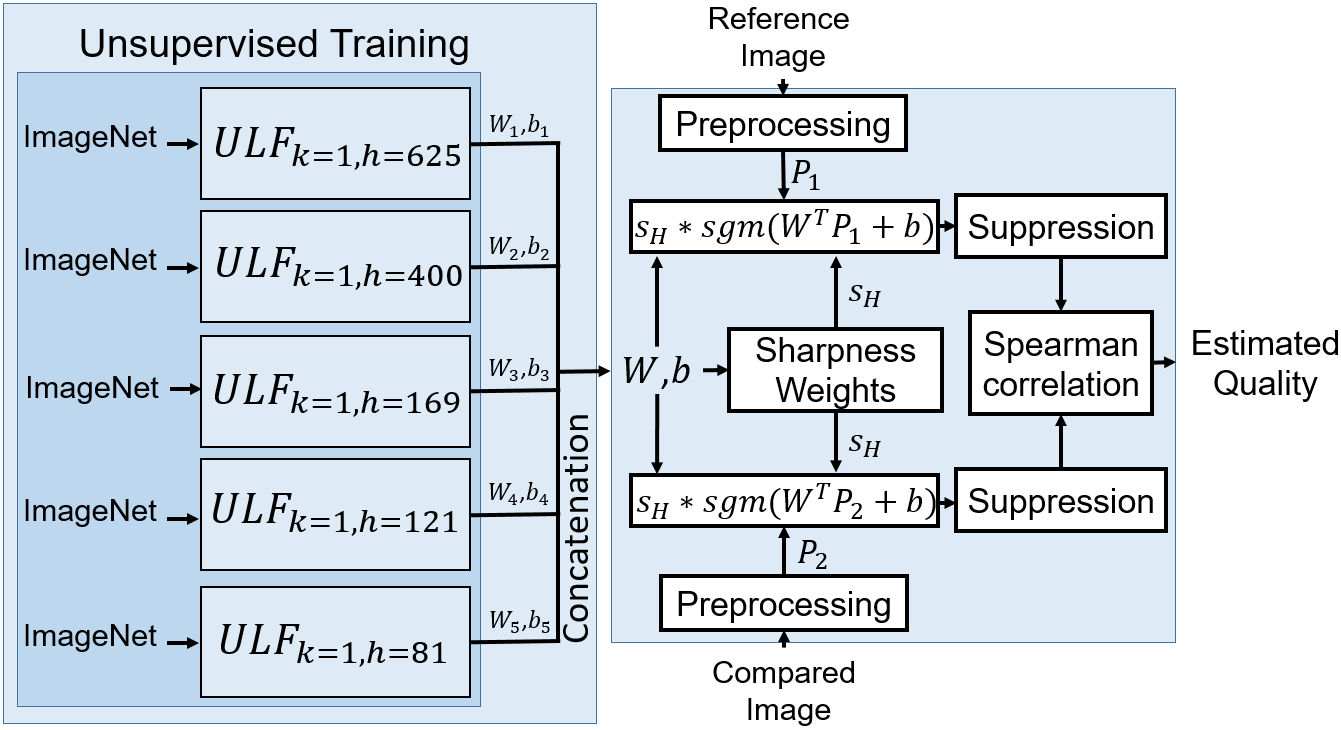}
		\caption{Multi-model and sharpness-weighted unsupervised image quality estimation.}
		\label{fig:MS-UNIQUE}
	\end{center}
	\vspace{-10.0mm}
\end{figure}
\vspace{-2mm}
\subsection{MS-UNIQUE: Multi-Model and Sharpness-Weighted Unsupervised Image Quality Estimation}
MS-UNIQUE was designed to have a wide configuration with independent networks modeling the same input using different number of neurons. Such a task is directly achieved by varying $h$ in the proposed framework, hence casting MS-UNIQUE as an instance of the framework. We use $5$ instances of $h$ spanning both undercomplete and overcomplete states. In MS-UNIQUE, we also classify the filters based on sharpness as either being activated by edge or color content. A detailed description of the algorithm is provided in \cite{prabhushankar2017}.
\vspace{-1.5mm}
\subsection{Validation}
A detailed validation of UNIQUE and MS-UNIQUE is provided in \cite{temel2016unique} and \cite{prabhushankar2017}. Table \ref{tab_results_databases} lists the results of state of the art estimators on LIVE \cite{sheikh2006statistical} and TID 2013 \cite{ponomarenko2015image} databases. The estimated scores from each method is validated against the subjective scores. Validation is performed utilizing metrics that analyze accuracy (RMSE), consistency (Outlier Ratio), linearity (Pearson correlation), and monotonic behavior (Spearman correlation). It is evident that both UNIQUE and MS-UNIQUE are among the top performing estimators. 
\begin{table}[htbp!]
	\scriptsize
	\centering
	\caption{Performance of image quality estimators.}
	\label{tab_results_databases}
	\begin{tabular}{|l|l|l|l|l|l|l|}
		\hline
		
		\multirow{3}{*}{{\bf Methods}}  \multirow{2}{*} &{\bf PSNR } &{\bf MS }  &{\bf SR }   &\multirow{2}{*}{{\bf FSIMc }}  \multirow{2}{*} &{{\bf UNI }}  &{\bf MS- } \\
		&\textbf{HMA} &{\bf SSIM} &{\bf SIM } & &{\bf QUE} &{\bf UNIQUE}
		\\ 
		&\cite{ponomarenko2011modified} & \cite{wang2004image}  & \cite{zhang2012sr} &\cite{zhang2011fsim} & \cite{temel2016unique} &\cite{prabhushankar2017} \\
		\hline 
		
		& \multicolumn{6}{c|}{\textbf{Outlier Ratio}}                                                                                                                                                                        
		
		\\ \hline
		\textbf{TID13}   
		&0.670   & 0.697   & \bf 0.632    &  0.727 &    0.640 &    \bf 0.611 

		\\ \hline

		& \multicolumn{6}{c|}{\textbf{Root Mean Square Error}}                                                                                                                                                                        \\ \hline
		\textbf{LIVE}  & \bf 6.58   & 7.43     &  7.54  &  7.20 &  6.76 &   \bf 6.61 \\ 
	
		\textbf{TID13}     &0.69   & 0.68   &   0.61 &   0.68  &  \bf 0.60  &  \bf 0.57
		\\ \hline              
		
		& \multicolumn{6}{c|}{\textbf{Pearson Correlation Coefficient}}                                                                                                                                                                        \\ \hline
		
		\multirow{1}{*}{{\bf LIVE}} & \bf 0.958   &  0.946    & 0.945  &0.950  &0.956  & \bf 0.958 \\
		
		\multirow{1}{*}{{\bf TID13}}&0.827   & 0.832    &0.866   &0.832  & \bf 0.870  & \bf 0.884 \\ \hline                         
		
		\textbf{}      & \multicolumn{6}{c|}{\textbf{Spearman Correlation Coefficient}}                                                                                                                                                                        \\ \hline
		
		\multirow{1}{*}{{\bf LIVE}} &0.944  &0.951  & \bf 0.955  & \bf 0.959 & 0.952 & 0.949 \\ 

		\multirow{1}{*}{{\bf TID13}} & 0.817 & 0.785 & 0.807  &0.851 & \bf 0.860 & \bf 0.870 \\ \hline 
	
	\end{tabular}
\end{table}
\vspace{-5mm}
\section{Texture Retrieval}
\label{sec:Texture}
The aim of texture retrieval is to detect images that are sampled from the same repeating texture as the given query image. Algorithms designed for recognizing and retrieving textures need to filter the entire contents of an image rather than non-overlapping patches of it. Hence, to perform a global filtering operation, we need to design a pyramidal structure of filter banks, which can be mimicked by stacking multiple hidden layers during training.
\vspace{-1.5mm}
\subsection{Training Filter Sets}
The block diagram in Fig. \ref{fig:Texture} gives an overview of the filter set generation module. We divide the filtering process into two phases based on the data characteristics we wish to capture - color based filtering and structure based filtering. For color based filtering, we train $400$ filters with $k=0$. Thus, the network is limited to learn only the first order statistics from unwhitened data. The filters are visualized in Fig. \ref{fig:Visualization Covariane Matrices}(d). 
For capturing structure, we process the entire image by building multiple layers of filter sets. Each layer is a self contained filter set, acting on a specific subregion. Following the illustration provided on the right side of Fig. \ref{fig:Texture}, we train filters to act on $8\times8\times3$ $P_2$ subregions. The responses of nine $P_2$ patches are concatenated to process $24\times24\times3$ $P_3$ subregions. The responses of adjoining nine $P_3$ patches are passed through a max pooling layer to obtain the maximum $64$ values from each $P_3$ subregion. Pooling is required to limit the input dimensions, since it is challenging for an autoencoder to extract features from high dimensional data. Moreover, pooling provides translation invariance. The pooled feature responses are concatenated to obtain a response vector that includes values from every part of an image. A final filter set is trained to transform these responses to obtain the final feature vector.
\vspace{-1.5mm}
\subsection{Retrieval}
\begin{figure}[htbp!]
	\begin{center}
		\noindent
		\includegraphics[width=1\linewidth]{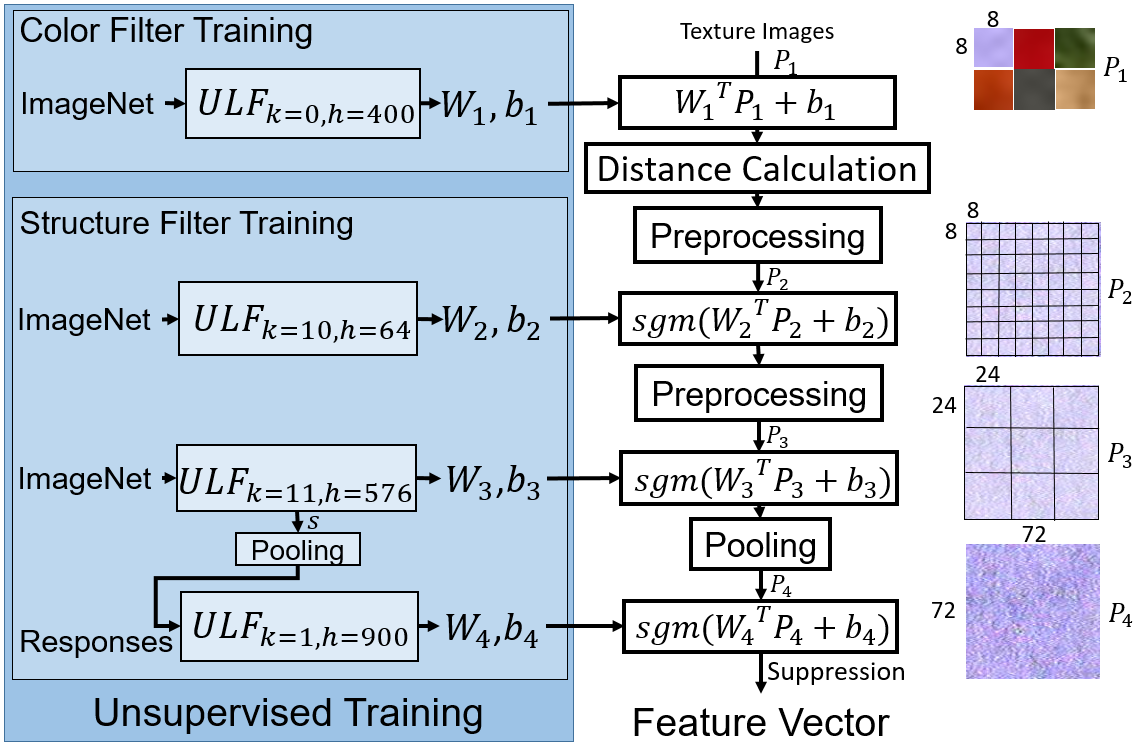}
		\caption{Texture retrieval through unsupervised learning framework.}
		\label{fig:Texture}
	\end{center}
\end{figure}

The texture images in the database are first resized to get patches $P_1$ of dimensions $8\times8\times3$, as shown on the top right of Fig. \ref{fig:Texture}. These low resolution images, essentially represent the mean color. Without any preprocessing, all the images are filtered using the generated color filter sets. The images with relatively similar colors to the query image are selected and examined for structure. Each texture image is resized to dimensions of $72\times72\times3$. The block diagram in Fig. \ref{fig:Texture} shows the progression of spatial filtering to obtain a final feature vector that encompasses the entire image. Once all feature vectors are obtained, Spearman correlation is calculated between all features in the database. Based on the similarity coefficients, similar texture images to the query image are retrieved.
\vspace{-10.0mm}
\subsection{Validation}
We test the retrieval process on the CUReT database \cite{dana1999curet}. Non overlapping patches of size $128\times128\times3$ were extracted from all images with viewing condition number $55$ as detailed by the authors in \cite{alfarraj2016content}. There are $61$ different texture classes, each of which have $3$ samples. To quantify the results of the proposed method, we use three standard retrieval metrics including Precision at one (P@1), Mean Reciprocal Rank (MRR), and Mean Average Precision (MAP). All of these metrics produce results in the range $[0,1]$ with $1$ being the ideal score. The results are presented in Table \ref{table:Texture Performance}. The proposed method is measured against commonly utilized and state of the art techniques. It can be observed that the proposed method is always among the top two performing metrics, second only to STSIM-1 \cite{ojala2002multiresolution}, which is a handcrafted technique designed to measure texture similarity. 
\vspace{-2.0mm}
\begin{table}[ht]
	\caption{Performance Validation on CUReT} 
	\small
	\centering 
	\begin{tabular}{c c c c } 
		\hline\hline 
		Method & P@1 & MRR & MAP \\ [0.5ex] 
		\hline 
		
		S-SSIM \cite{wang2004image} & 0.0546 & 0.0952 & 0.0935 \\
		CW-SSIM \cite{sampat2009complex} & 0.1530 & 0.2612 & 0.1925 \\
		\bf STSIM-1 \cite{ojala2002multiresolution} & \bf 0.9071 & \bf 0.9447 & \bf 0.9048 \\
		STSIM-2 \cite{ojala2002multiresolution} & 0.8852 & 0.9248 & 0.8500  \\
		STSIM-M \cite{ojala2002multiresolution} & 0.8798 & 0.9170 & 0.8516  \\
		\bf Proposed & \bf 0.9126 & \bf 0.9371 & \bf 0.8922  \\ [1ex] 
		\hline 
	\end{tabular}
	\label{table:Texture Performance} 
\end{table}
\vspace{-7.0mm}
\subsection{Robustness}
We examine the proposed method's robustness against noise. Texture retrieval is performed after adding random noise to CUReT database, with standard deviation $\sigma$ in the range $[0,100]$. The MAP results of top metrics are presented in Table \ref{table:Texture Robustness}.
\begin{table}[ht]
	\caption{MAP values for random Gaussian noise with zero mean and standard deviation $\sigma$} 
	\scriptsize
	\centering 
	\begin{tabular}{c c c c c c} 
		\hline\hline 
		$\sigma$ & CW-SSIM & STSIM-1 & STSIM-2 & STSIM-M & Proposed \\ [0.5ex] 
		\hline 
		
		5 & 0.1939 & \bf 0.9017 & 0.8433 & 0.8579 & 0.8931 \\
		25 & 0.1700 & 0.8576 & 0.7478 & 0.8033 & \bf 0.8600 \\
		50 & 0.1185 & 0.7450 & 0.5839 & 0.6990 & \bf 0.8535 \\
		75 & 0.1204 & 0.6825 & 0.4460 & 0.6218 & \bf 0.8351 \\
		100 & 0.0876 & 0.5465 & 0.3487 & 0.4826 & \bf 0.8219 \\ [1ex] 
		\hline 
	\end{tabular}
	\label{table:Texture Robustness} 
\end{table}
The results clearly indicate that the drop off of MAP values for the proposed method is lower than the compared techniques. Robustness of the proposed method is rooted in the expectation that a higher level representation should be stable under corruptions to the input. Using the $k$ extension to the standard ZCA algorithm decorrelates adjoining pixel values in the natural images, hence disrupting local structure in the data - a behavior equivalent to adding noise. The proposed method has been trained on decorrelated data and has learnt to consider only the underlying principal components that constitute an image ignoring the lower order statistics.
\vspace{-1.5mm}
\section{Conclusion}
\label{sec:Conclusion}
\vspace{-1.5mm}
In this paper, we proposed the Unsupervised Learning Framework (ULF) that is based on a data driven approach, to solve challenges which lack sufficient domain-specific data and annotations. An extension to the classical ZCA algorithm was proposed that would, in practice, orthogonalize the input vectors thereby eliminating all the lower order natural scene statistics. Orthogonalization allows the linear decoder to learn higher order characterizations from the data. We modified the number of neurons in the linear decoder to obtain multiple filter sets modeling the same input and used them in parallel. We demonstrated the use of these filter sets on image quality assessment (IQA) and texture retrieval. In IQA, we showed that the already proposed estimators, UNIQUE and MS-UNIQUE are instances of the framework and that the filter sets spanned a response space wherein perceptual dissimilarity could be measured. We also showed that similarity is quantifiable in the same response space, by retrieving texture images. We established the robustness of the proposed method towards noisy input data. In conclusion, we illustrated utilizing unsupervised learning to generate robust and adaptive filter sets that can be used in various image processing applications.

\begin{thebibliography}{10}

\bibitem{Goodfellow-et-al-2016}
I.~Goodfellow, Y.~Bengio, and A.~Courville,
\newblock {\em Deep Learning},
\newblock MIT Press, 2016.

\bibitem{vincent2010stacked}
P.~Vincent, H.~Larochelle, I.~Lajoie, Y.~Bengio, and P.-A. Manzagol,
\newblock ``Stacked denoising autoencoders: Learning useful representations in
  a deep network with a local denoising criterion,''
\newblock {\em Journal of Machine Learning Research}, vol. 11, no. Dec, pp.
  3371--3408, 2010.

\bibitem{bell1997independent}
A.~J. Bell and T.~J. Sejnowski,
\newblock ``The “independent components” of natural scenes are edge
  filters,''
\newblock {\em Vision research}, vol. 37, no. 23, pp. 3327--3338, 1997.

\bibitem{krizhevsky2009learning}
A.~Krizhevsky and G.~Hinton,
\newblock ``Learning multiple layers of features from tiny images,''
\newblock 2009.

\bibitem{ng2011sparse}
A.~Ng,
\newblock ``Sparse autoencoder,''
\newblock {\em CS294A Lecture notes}, vol. 72, no. 2011, pp. 1--19, 2011.

\bibitem{ng2012ufldl}
A.~Ng, J.~Ngiam, C.~Y. Foo, Y.~Mai, and C.~Suen,
\newblock ``Ufldl tutorial,'' 2012.

\bibitem{russakovsky2015imagenet}
O.~Russakovsky, J.~Deng, H.~Su, J.~Krause, S.~Satheesh, S.~Ma, Z.~Huang,
  A.~Karpathy, A.~Khosla, M.~Bernstein, et~al.,
\newblock ``Imagenet large scale visual recognition challenge,''
\newblock {\em International Journal of Computer Vision}, vol. 115, no. 3, pp.
  211--252, 2015.

\bibitem{charrier2012}
C.~Charrier, O.~L{\'e}zoray, and G.~Lebrun,
\newblock ``Machine learning to design full-reference image quality assessment
  algorithm,''
\newblock {\em Signal Processing: Image Communication}, vol. 27, no. 3, pp.
  209--219, 2012.

\bibitem{tang2014blind}
H.~Tang, N.~Joshi, and A.~Kapoor,
\newblock ``Blind image quality assessment using semi-supervised rectifier
  networks,''
\newblock in {\em Proceedings of the IEEE Conference on Computer Vision and
  Pattern Recognition}, 2014, pp. 2877--2884.

\bibitem{ye2013real}
P.~Ye, J.~Kumar, L.~Kang, and D.~Doermann,
\newblock ``Real-time no-reference image quality assessment based on filter
  learning,''
\newblock in {\em Proceedings of the IEEE Conference on Computer Vision and
  Pattern Recognition}, 2013, pp. 987--994.

\bibitem{chang2011sparse}
H.~Chang and M.~Wang,
\newblock ``Sparse correlation coefficient for objective image quality
  assessment,''
\newblock {\em Signal processing: Image communication}, vol. 26, no. 10, pp.
  577--588, 2011.

\bibitem{temel2016unique}
D.~Temel, M.~Prabhushankar, and G.~AlRegib,
\newblock ``Unique: Unsupervised image quality estimation,''
\newblock {\em IEEE Signal Processing Letters}, vol. 23, no. 10, pp.
  1414--1418, 2016.

\bibitem{prabhushankar2017}
M.~Prabhushankar, D.~Temel, and G.~AlRegib,
\newblock ``Ms-unique: Multi-model and sharpness-weighted unsupervised image
  quality estimation,''
\newblock {\em Electronic Imaging}, vol. 2017, no. 12, 2017.

\bibitem{sheikh2006statistical}
H.~R. Sheikh, M.~F. Sabir, and A.~C. Bovik,
\newblock ``A statistical evaluation of recent full reference image quality
  assessment algorithms,''
\newblock {\em IEEE Transactions on image processing}, vol. 15, no. 11, pp.
  3440--3451, 2006.

\bibitem{ponomarenko2015image}
N.~Ponomarenko, L.~Jin, O.~Ieremeiev, V.~Lukin, K.~Egiazarian, J.~Astola,
  B.~Vozel, K.~Chehdi, M.~Carli, F.~Battisti, et~al.,
\newblock ``Image database tid2013: Peculiarities, results and perspectives,''
\newblock {\em Signal Processing: Image Communication}, vol. 30, pp. 57--77,
  2015.

\bibitem{ponomarenko2011modified}
N.~Ponomarenko, O.~Ieremeiev, V.~Lukin, K.~Egiazarian, and M.~Carli,
\newblock ``Modified image visual quality metrics for contrast change and mean
  shift accounting,''
\newblock in {\em CAD Systems in Microelectronics (CADSM), 2011 11th
  International Conference The Experience of Designing and Application of}.
  IEEE, 2011, pp. 305--311.

\bibitem{wang2004image}
Z.~Wang, A.~C. Bovik, H.~R. Sheikh, and E.~P. Simoncelli,
\newblock ``Image quality assessment: from error visibility to structural
  similarity,''
\newblock {\em IEEE transactions on image processing}, vol. 13, no. 4, pp.
  600--612, 2004.

\bibitem{zhang2012sr}
L.~Zhang and H.~Li,
\newblock ``Sr-sim: A fast and high performance iqa index based on spectral
  residual,''
\newblock in {\em Image Processing (ICIP), 2012 19th IEEE International
  Conference on}. IEEE, 2012, pp. 1473--1476.

\bibitem{zhang2011fsim}
L.~Zhang, L.~Zhang, X.~Mou, and D.~Zhang,
\newblock ``Fsim: A feature similarity index for image quality assessment,''
\newblock {\em IEEE transactions on Image Processing}, vol. 20, no. 8, pp.
  2378--2386, 2011.

\bibitem{dana1999curet}
K.~J. Dana, B.~van Ginneken, S.~Nayar, and J.~Koenderink,
\newblock ``Curet: Columbia-utrecht reflectance and texture database,'' 1999.

\bibitem{alfarraj2016content}
M.~Alfarraj, Y.~Alaudah, and G.~AlRegib,
\newblock ``Content-adaptive non-parametric texture similarity measure,''
\newblock in {\em Multimedia Signal Processing (MMSP), 2016 IEEE 18th
  International Workshop on}. IEEE, 2016, pp. 1--6.

\bibitem{ojala2002multiresolution}
T.~Ojala, M.~Pietikainen, and T.~Maenpaa,
\newblock ``Multiresolution gray-scale and rotation invariant texture
  classification with local binary patterns,''
\newblock {\em IEEE Transactions on pattern analysis and machine intelligence},
  vol. 24, no. 7, pp. 971--987, 2002.

\bibitem{sampat2009complex}
M.~P. Sampat, Z.~Wang, S.~Gupta, A.~C. Bovik, and M.~K. Markey,
\newblock ``Complex wavelet structural similarity: A new image similarity
  index,''
\newblock {\em IEEE transactions on image processing}, vol. 18, no. 11, pp.
  2385--2401, 2009.

\end{thebibliography}

\end{document}